\documentclass[preprint,showpacs,preprintnumbers,amsmath,amssymb,showkeys]{revtex4}

\usepackage[unicode,bookmarks,bookmarksopen,bookmarksopenlevel=0,colorlinks,%
pageanchor=1,breaklinks=1,linkcolor=blue,citecolor=blue,urlcolor=red,hypertex]{hyperref}

\usepackage{graphicx}% Include figure files
\usepackage{dcolumn}% Align table columns on decimal point
\usepackage{bm}% bold math

\usepackage{float}

\begin{document}

\preprint{APS/123-QED}

\title{Mathematical Model of a pH-gradient Creation at~Isoelectrofocusing.\\ Part II. Numerical Solution of the Stationary Problem.}

\author{L.\,V.~Sakharova}
\email{l_sakharova@mail.ru}
\affiliation{%
Institute of Water Transport \\ Rostov-on-Don, Russia
}%

\author{E.\,V.~Shiryaeva}
\email{shir@math.sfedu.ru}

\author{M.\,Yu.~Zhukov}%
 \email{myuzhukov@gmail.com}
\affiliation{%
Southern Federal University\\ Rostov-on-Don, Russia
}%

\date{\today}% It is always \today, today,

\begin{abstract}
The mathematical model describing the natural $\textrm{pH}$-gradient arising under the action of an electric field in an aqueous solution of ampholytes (amino acids)  is constructed and investigated. This paper is the second part of the series papers \cite{Part1,Part3,Part4} that are devoted to $\textrm{pH}$-gradient creation problem. We present the numerical solution of the stationary problem. The equations system has a small parameter at higher derivatives and the turning points, so called stiff problem. To solve this problem numerically we use the shooting method: transformation of the boundary value problem to the Cauchy problem.  At large voltage or electric current density we compare the numerical solution with weak solution presented in \cite{Part1}.

\end{abstract}


%%%%    82.45.-h Electrochemistry and electrophoresis;  electrochemistry, 82.45.-h
%%%%    87.15.Tt Electrophoresis
%%%%    82.80.Yc pH measurement
%%%%    02.60.-x Numerical methods (mathematics)
%%%%    82.45.Tv, 87.15.Tt   Electrophoresis biomolecule,
%%%%    87.50.ch  Biological systems electrophoresis,
%%%%    Polyelectrolytes, 82.35.Rs
%%%%    in electrochemistry, 82.45.Wx
%%%%    Transport processes
%%%%    classical, 05.60.Cd
%Use showkeys class option if keyword
\maketitle

\section{Introduction}\label{ZhS-01}

From the mathematical point of view, the modeling of stationary natural $\textrm {pH}$-gradients problem  is reduced to the solution of the ODE's equations for distribution of concentration, some algebraic constrain, and integral conditions. At large intensity of the electric field (or large density of an electric current) the system of the equations is stiff: ODE's have the small parameter at the highest  derivatives and have the turning points. Numerical integration of this problem becomes complicated also that solutions for separate concentration are focused in some regions of the integration interval and quickly exponential decrease out of these regions.

In previous paper \cite{Part1} we showed that at limiting case the solution of the problem tends to a weak solution when the $\textrm {pH}$-gradient has the piecewise constant profile and concentrations  have almost rectangle profile. In this paper we confirm these results with the help of numerical solution.  We also present the results of calculation for real mixtures of amino acids.

The paper is organized as follows. In Sec.~\ref{App:ZhS-2n} the basic equations of the stationary electrophoresis problem obtained in \cite{Part1} are presented.
The numerical methods and numerical experiments are described in Sec.~\ref{App:ZhS-2}, where we present the results of calculation for real mixtures of ampholites and compare results with weak piecewise constant solution of the problem obtained in \cite{Part1}.

%%%%% begin zhuk %%%%%%%

%\appendix

\section{Basic equations}\label{App:ZhS-2n}

In detail the equations system describing the creation of $\textrm {pH}$-gradient for stationary case is given in \cite{Part1} (see also \cite{ZhukovBabskiyYudovich,BabZhukYudE,MosherSavilleThorman,ZhRStoy2001}).
In dimensionless variables this system has the following form
\begin{equation}\label{ZhS-eq7}
  \frac{1}{\lambda}\frac{da_k}{dx} =
  \frac{a_k\theta_k(\psi)\sum\limits_{i=1}^n a_i\theta'_i(\psi)}
  {\sigma\sum\limits_{i=1}^n  a_i \left( \theta^2_i(\psi) + \theta'_i(\psi)\right)}, \quad  k=1,...,n,
  \quad  0 \leqslant x \leqslant L,\quad \lambda=\frac{j_0}{\varepsilon},
\end{equation}
\begin{equation}\label{ZhS-eq8}
  \sum\limits_{k=1}^n a_k \theta_k(\psi) = 0,
\end{equation}
\begin{equation}\label{ZhS-eq9}
   \int\limits_0^L a_k(x)\,dx = M_k,
\end{equation}
\begin{equation}\label{ZhS-eq10}
 \sigma =\sum_{i=1}^n \mu_i a_i \theta'_i(\psi), \quad \theta_i(\psi)=\frac{\varphi'_i(\psi)}{\varphi_i(\psi)},
 \quad \varphi_i(\psi)=\cosh(\psi-\psi_i)+\delta_i,
\end{equation}
where $a_k$ are the analytical concentrations of the mixture component, $\psi$ is the acidity function of the mixture, $\theta_{i}(\psi)$ is the specific molar charge of the component, $\sigma$ is the specific molar conductivity of the mixture, $L$  is the length of the electrophoretic chamber, $M_k$ is the quantity  of the concentration  $a_k$  on the interval $[0,L]$, $\delta_i>0$ is the dimensionless parameter that characterize component, $\psi_i$ is the isoelectric point (electrophoretic mobility is equal to zero at $\psi=\psi_i$),  $\varphi_i(\psi)$ is some auxiliary function, $\varepsilon$ is characteristic diffusion coefficient, $j_0$ is constant current density.

The system (\ref{ZhS-eq7})--(\ref{ZhS-eq10}) has integral which one can get by the summation of all equations (\ref{ZhS-eq7}) and taking into account (\ref{ZhS-eq8}):
\begin{equation}\label{ZhS-eq11}
 \sum_{i=1}^n a_i =a_0\equiv L^{-1}\sum\limits_{i=1}^n M_i,
\end{equation}
where the constant $a_0$ is defined by (\ref{ZhS-eq9}).

Note, that the solution of (\ref{ZhS-eq7})--(\ref{ZhS-eq11}) for large values of the parameter $\lambda$ involves difficulties due to the presence of a small parameter at highest derivatives and the turning points at $\psi=\psi_i$. Additional difficulties arise from the fact that the location of the turning points in the interval $[0,L]$ depends on the acidity function $\psi(x)$, \emph{i.\,e.} the solution of the problem.

\subsection{Potential difference}\label{App:ZhS-2.1}

The model (\ref{ZhS-eq7})--(\ref {ZhS-eq10}) describes IEF process at constant electric current density $j=j_0$. The potential difference across the electrophoretic chamber  (or voltage $U$) can be calculate with the help of formulae
\begin{equation}\label{ZhSeq-B7}
  j=j_0=\sigma(E-\varepsilon\psi_x),
\end{equation}
\begin{equation*}
  U=\int\limits_0^L E\,dx=\frac{j_0}{\lambda}(\psi(L)-\psi(0))+U_0, \quad U_0=j_0\int\limits_0^L \frac{1}{\sigma}\,dx.
\end{equation*}

In the case of very large $\lambda$ one can calculate the approximation value of $U_0$ using formula
\begin{equation}\label{ZhSeq-B8}
  U_0=j_0\int\limits_0^L \frac{1}{\sigma}\,dx\approx j_0 \sum\limits_{k=1}^{n}\frac{H_k}{\mu_k\theta'_k(\psi_k)a_0}=
  j_0 \sum\limits_{k=1}^{n}\frac{M_k(1+\delta_k)}{\mu_k a_0^2}.
\end{equation}

\section{Numerical solution}\label{App:ZhS-2}

For numerical integration of the original problem (\ref{ZhS-eq7})--(\ref {ZhS-eq10}) we use the transformation describing in \cite{SakhVladZhuk} (see also \cite{Averkov,SakhSKNC,SakhOrel}):
\begin{equation}\label{ZhSeq-B1}
  a_k(x)=c_k(x) \varphi_k(\psi(x)), \quad m_k(x)=\int\limits_0^x c_k(y) \varphi_k(\psi(y))\,dy.
\end{equation}
For obtaining $c_k$, $m_k$ we have the boundary problem instead the problem with integral conditions (\ref{ZhS-eq9}):
\begin{equation}\label{ZhSeq-B2}
  \frac{dc_k}{dx} =
  \frac{\lambda \theta_k(\psi) c_k}{\sigma}, \quad
  \frac{dm_k}{dx} = c_k \varphi_k(\psi),  \quad  k=1,...,n,
\end{equation}
\begin{equation}\label{ZhSeq-B3}
  m_k(0)=0, \quad m_k(L)=M_k, \quad  k=1,...,n,
\end{equation}
where
\begin{equation}\label{ZhSeq-B4}
  \psi=\frac12\ln\frac{\displaystyle\sum\limits_{i=1}^n c_i \exp(\psi_i)}{\displaystyle\sum\limits_{i=1}^n c_i \exp(-\psi_i)},\quad
  \sigma =\sum_{i=1}^n \mu_i c_i \varphi_i(\psi)\theta'_i(\psi).
\end{equation}
To solve the problem (\ref{ZhSeq-B2})--(\ref {ZhSeq-B4}) the shooting method simultaneously with method of moving parameters is used. At $\lambda=0$ the solution of the boundary problem is:
\begin{equation}\label{ZhSeq-B5}
  c_k^0=\frac{M_k}{L\varphi_k(\psi^0)},
\end{equation}
where $\psi^0$ is root of equation
\begin{equation}\label{ZhSeq-B6}
  \Phi(\psi^0) \equiv \sum_{i=1}^n M_k\theta_k(\psi^0)=0.
\end{equation}
Taking into account that
\begin{equation*}
  \Phi'(\psi^0)>0,\quad \Phi(\psi_n)<0, \quad  \Phi(\psi_1)>0, \quad \psi_n<\psi_{n-1}<\cdots<\psi_2<\psi_1
\end{equation*}
we can solve the equation (\ref{ZhSeq-B6}) with the help, for instance, of the bisection method.

To obtain the solution for some $\lambda=\lambda_0$ we use the initial approximation  $c_k^0$, $\psi^0$. Changing the $\lambda$ from $\lambda=0$ by step $\Delta\lambda$ to $\lambda=\lambda_0$ we get solution of the problem (\ref{ZhSeq-B2})--(\ref {ZhSeq-B4}) at $\lambda=\lambda_0$.

\subsection{Numerical experiment}\label{App:ZhS-2.2}

We demonstrate the functionality of the numerical algorithm  for the solving of practical IEF problem. The characteristic values of the parameters are given in the Tab.~\ref{tab:table3}. For specificity we assume that the electrophoretic chamber is the capillary of the radius $r_*$ and the sectional area $S_*$.

The dimensional current density $j_*$, current $I_*$, and voltage $U_*$ are connected with dimensionless parameter $\lambda$ by formulae
\begin{equation}\label{ZhSeq-B9}
  j_*=\lambda \frac{R_*T_*C_*\mu_*}{L_*}\approx 0.025\,\lambda\,\,(\textrm{A}\cdot\textrm{m}^{-2}),
\end{equation}
\begin{equation*}
  I_*=j_*S_*\approx 0.25\cdot10^{-6}\,\lambda (\,\,\textrm{A}),
\end{equation*}

\begin{equation*}
 U_*=UE_*L_*=\frac{R_*T_*}{F_*}\frac{U\lambda}{j_0}\approx 0.025 \frac{U\lambda}{j_0}\,\,(\textrm{V}).
\end{equation*}

\begin{table}[H]
\caption{\label{tab:table3} Dimensional parameters}
\begin{ruledtabular}
\begin{tabular}{lll}
$L_*$ & $0.1\,\textrm{m}$ & Length \\
\hline
$C_*$ & $100\,\,\textrm{mol}/\textrm{m}^3=0.1\,\,\textrm{mol}/\textrm{litr} $ & Concentration \\
\hline
%%$E_*$ & $100\,\,\textrm{V}/\textrm{m}$ & Intensity of electric field \\
%%\hline
$\mu_*$ & $10^{-8}\,\,\textrm{m}^2/(\textrm{V}\cdot\textrm{s})$ & Electrophoretic mobility \\
\hline
$F_*$ & $96485.34\,\,\textrm{C}/\textrm{mol}$ & Faraday's number \\
\hline
$R_*$ & $8.314462\,\,\textrm{J}/(\textrm{mol}\cdot\textrm{K})$ & Universal gas constant \\
\hline
$T_*$ & $293\,\,\textrm{K}$ & Temperature \\
\hline
$\sigma_*=F_*C_*\mu_*$ & $0.09648534\,\textrm{C}/(\textrm{m}\cdot\textrm{V}\cdot\textrm{s})$ & Conductivity \\
\hline
$S_*$ & $ 10^{-5}\,\,\textrm{m}^2$ & Sectional area \\
\hline
$r_*$ & $ 0.0018\,\,\textrm{m}$ & Radius of capillary \\
\end{tabular}
\end{ruledtabular}
\end{table}

\subsubsection{Five-component mixture}\label{App:ZhS-2.2.1}

\begin{figure}[H]
\centering
\includegraphics[scale=0.85]{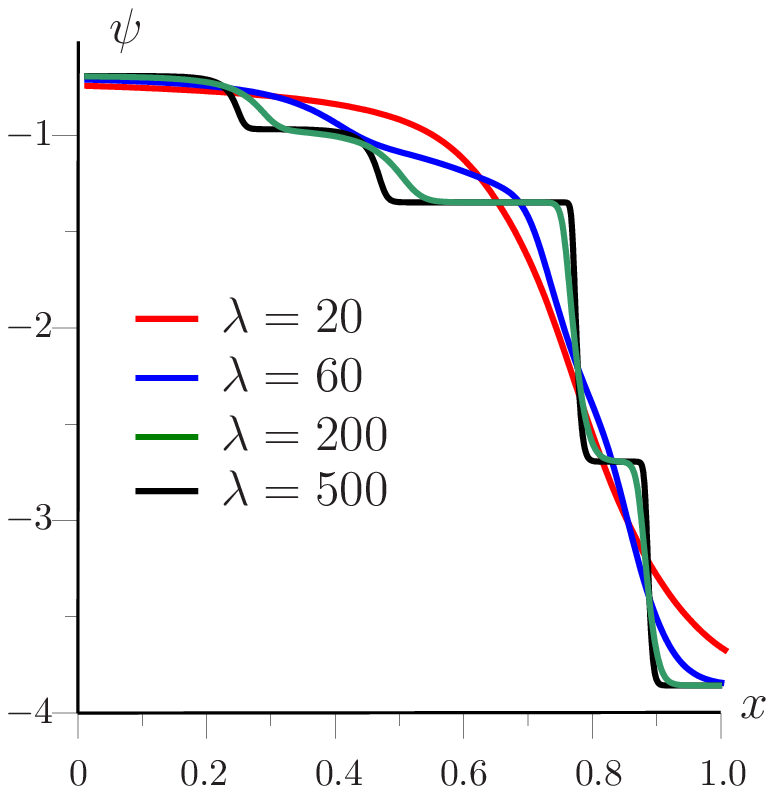}\quad
\includegraphics[scale=0.85]{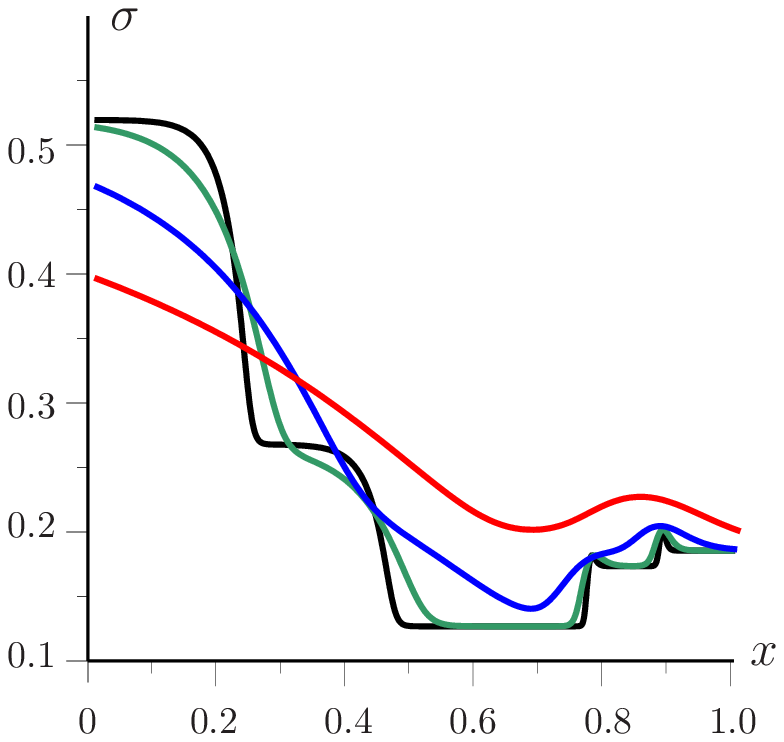}\\ [-4mm]
\caption{The distributions of the acidity $\psi(x)$ and the conductivity $\sigma(x)$ at
 $\lambda=20$,\\  $U_0=3.825$ ($I_*=4.959   \,\,\mu\textrm{A}$,  $U_*=1.852  \,\,\textrm{V}$);
 $\lambda=60$,  $U_0=4.413$ ($I_*=14.878  \,\,\mu\textrm{A}$,  $U_*=6.605  \,\,\textrm{V}$);
 $\lambda=200$, $U_0=4.897$ ($I_*=49.594  \,\,\mu\textrm{A}$,  $U_*=24.649 \,\,\textrm{V}$);
 $\lambda=500$, $U_0=5.035$ ($I_*=123.984 \,\,\mu\textrm{A}$,  $U_*= 63.484\,\,\textrm{V}$). See table~\ref{tab:table4}}
\label{Ris7}
\end{figure}

\begin{figure}[H]
\centering
\includegraphics[scale=0.8]{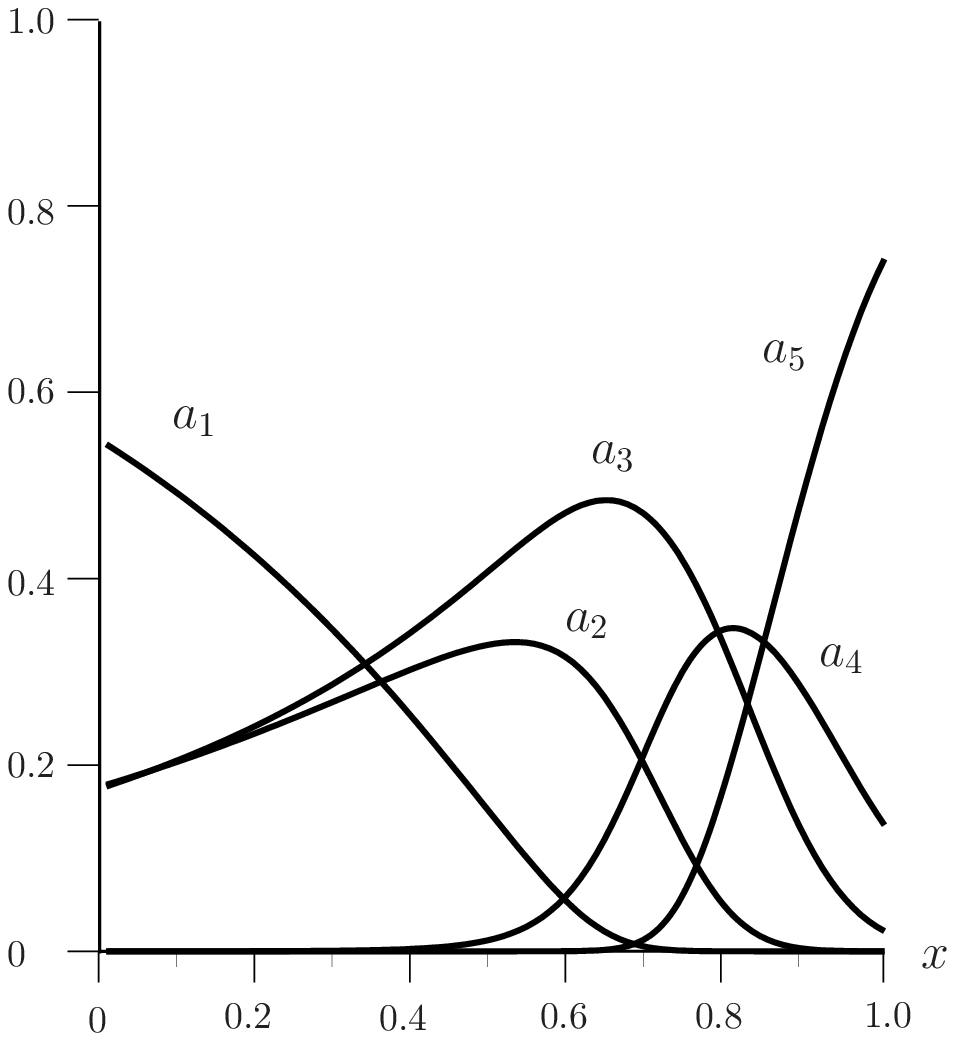}
\includegraphics[scale=0.8]{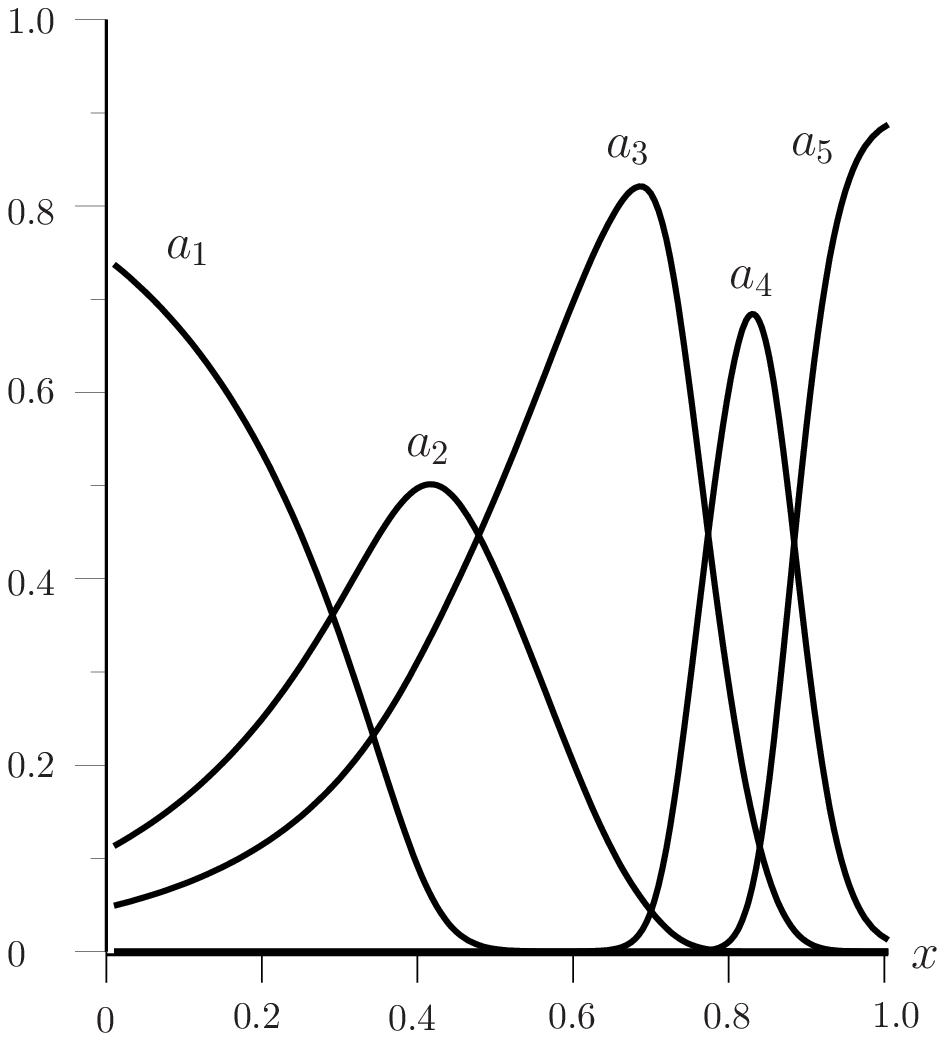}\\
\caption{The amino acid concentrations at $\lambda=20$ and $\lambda=60$. See table~\ref{tab:table4}}
\label{Ris8}
\end{figure}

\begin{figure}[H]
\centering
\includegraphics[scale=0.8]{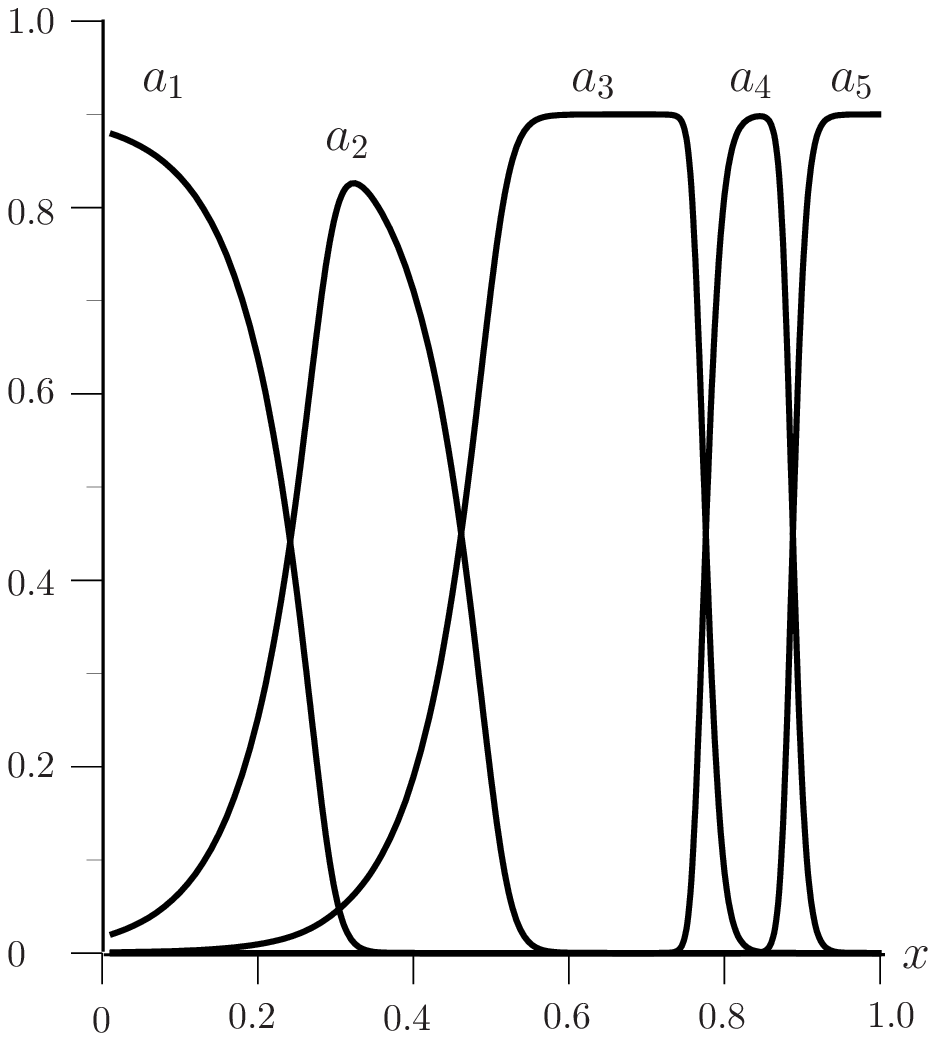}
\includegraphics[scale=0.8]{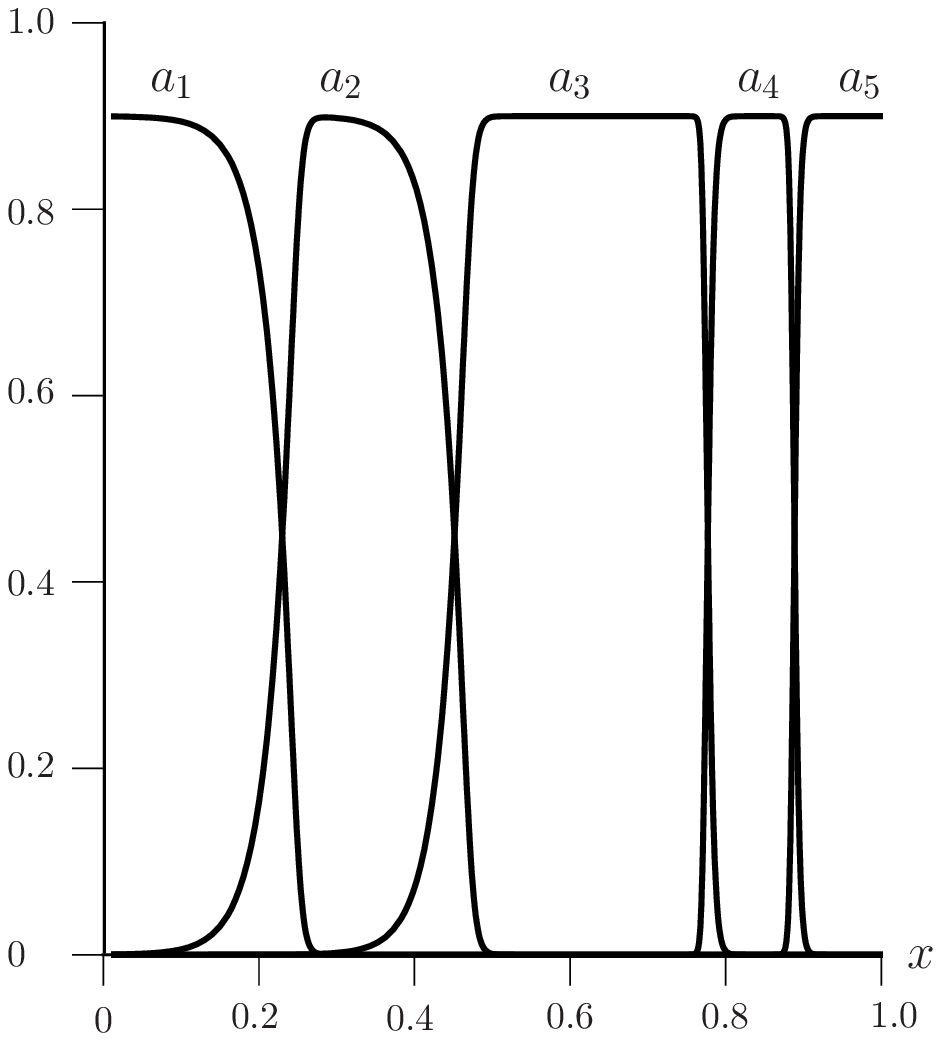}\\  [-3mm]
\caption{The amino acid concentrations at $\lambda=200$ and $\lambda=500$. See table~\ref{tab:table4}}
\label{Ris9}
\end{figure}

The Figs.~\ref{Ris7}--\ref{Ris9}  are demonstrated the numerical results for five-component mixture of the amino acids $\textrm{His-His}$, $\textrm{His-Gly}$, $\textrm{His}$, $\beta-\textrm{Ala-His}$, $\textrm{Tyr-Arg}$. In Tab.~\ref{tab:table4} the parameters of mixture components are given (see \cite{Righetti83}).

\begin{table}[H]
\caption{\label{tab:table4} Parameters of amino acids for five-component mixture (see \cite{Righetti83})}
\begin{ruledtabular}
\begin{tabular}{rccccccc}
 & $\textrm{pKb}_i$ & $\textrm{pKa}_i$ & $\textrm{pI}_i$ & $\psi_i$ & $\delta_i$  & $\mu_i$& $ M_i$\\
\hline
His-His         \quad ($a_1$) & $6.80$ & $7.80$ & $7.300$ & $-0.691$ & $1.58$   & $1.49$ & $0.2$ \\
His-Gly         \quad ($a_2$) & $6.27$ & $8.57$ & $7.420$ & $-0.967$ & $7.06$   & $2.40$ & $0.2$ \\
His             \quad ($a_3$) & $6.00$ & $9.17$ & $7.585$ & $-1.347$ & $19.23$  & $2.85$ & $0.3$ \\
$\beta$-Ala-His \quad ($a_4$) & $6.83$ & $9.51$ & $8.170$ & $-2.694$ & $10.94$  & $2.30$ & $0.1$ \\
Tyr-Arg         \quad ($a_5$) & $7.55$ & $9.80$ & $8.675$ & $-3.857$ & $6.67$   & $1.58$ & $0.1$ \\
\end{tabular}
\end{ruledtabular}
%\footnotetext[1]{$B_i=10^{-\textrm{pKb}_i}$.}
%\footnotetext[2]{$A_i=10^{-\textrm{pKa}_i}$.}
\end{table}

On Fig.~\ref{Ris7} the stationary distributions of the acidity $\psi(x)$ and the conductivity $\sigma(x)$ are shown at $\lambda=20$,
$\lambda=60$, $\lambda=200$, $\lambda=500$. Starting from $\lambda=200$ the functions $\psi(x)$ and  $\sigma(x)$ have almost the piecewise constant shape.

On Figs.~\ref{Ris8}, \ref{Ris9} the stationary distributions of amino acid are shown at $\lambda=20$,
$\lambda=60$, $\lambda=200$, $\lambda=500$.

\begin{figure}[H]
\centering
\includegraphics[scale=0.7]{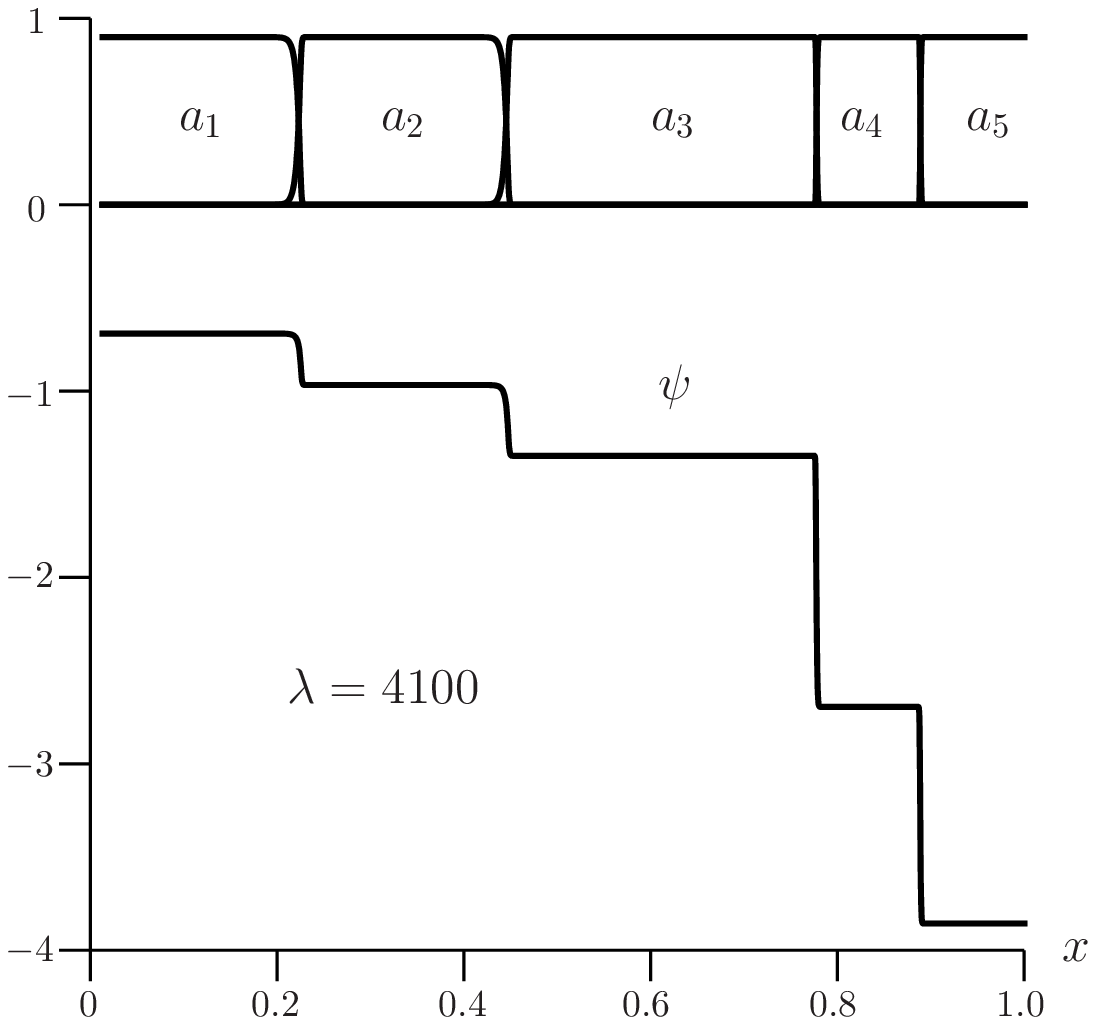}
\includegraphics[scale=0.7]{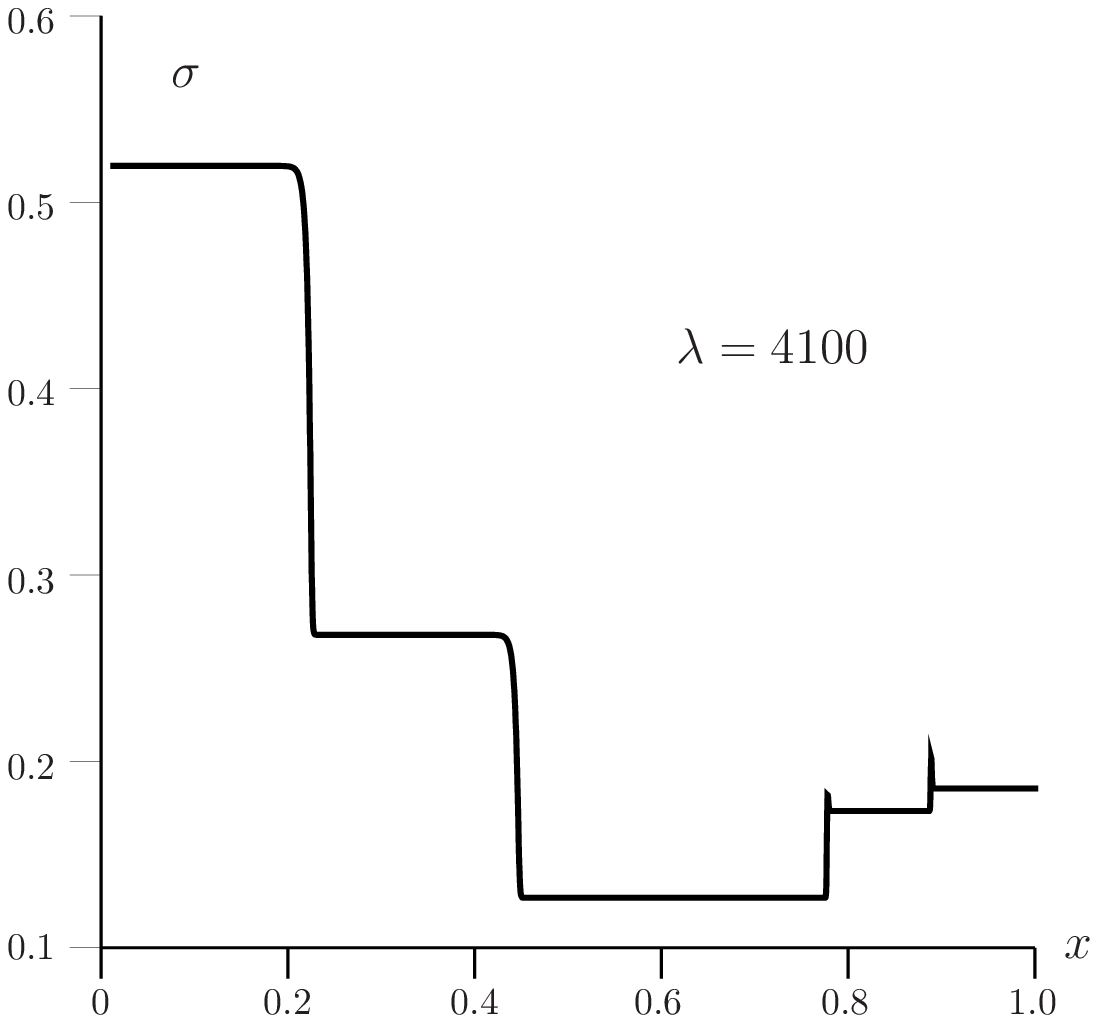}\\ [-3mm]
\caption{The distributions of the acidity $\psi(x)$, concentration $a_k(x)$, and the conductivity $\sigma(x)$ at
\\ $\lambda=4100$,  $U_0=5.115$ ($I_*=1016.671 \,\,\mu\textrm{A}$,  $U_*=529.425  \,\,\textrm{V}$). See table~\ref{tab:table4}}
\label{Ris10}
\end{figure}

The Fig.~\ref{Ris10} demonstrates the stationary distributions of the acidity $\psi(x)$, concentrations $a_1(x)$, $a_2(x)$, $a_3(x)$, $a_4(x)$, $a_5(x)$, and the conductivity $\sigma(x)$ at $\lambda=4100$.

This result have a good agreement to formulae (B3), (B4) obtained in \cite{Part1}. At $\lambda\to\infty$ the solution of the problem tend effectively to general piecewise constant solution
\begin{equation} \label{ZhSeq-E3}
a_k(x)= \left\{
\begin{array}{ll}
0, & x \leqslant \overline{x}_{k-1}, \\
a_0, & \overline{x}_{k-1} \leqslant x\leqslant \overline{x}_k, \\
0, & \overline{x}_k \leqslant x,
\end{array}
\right.
\quad
\psi(x)=\psi_k,\quad \overline{x}_{k-1} \leqslant x\leqslant \overline{x}_k, \\
\quad k=1,\dots,n,
\end{equation}
where
\begin{equation}\label{ZhSeq-E4}
   \overline{x}_0=0, \quad \overline{x}_k=\overline{x}_{k-1}+M_ka_0^{-1},  \quad k=1,\dots,n, \quad \overline{x}_n=L, \quad
   a_0=L^{-1}\sum\limits_{i=1}^n M_i
\end{equation}

\subsubsection{Separation samples in $\textrm{pH}$-gradient}\label{App:ZhS-2.2.2}

\begin{figure}[H]
\centering
\includegraphics[scale=0.85]{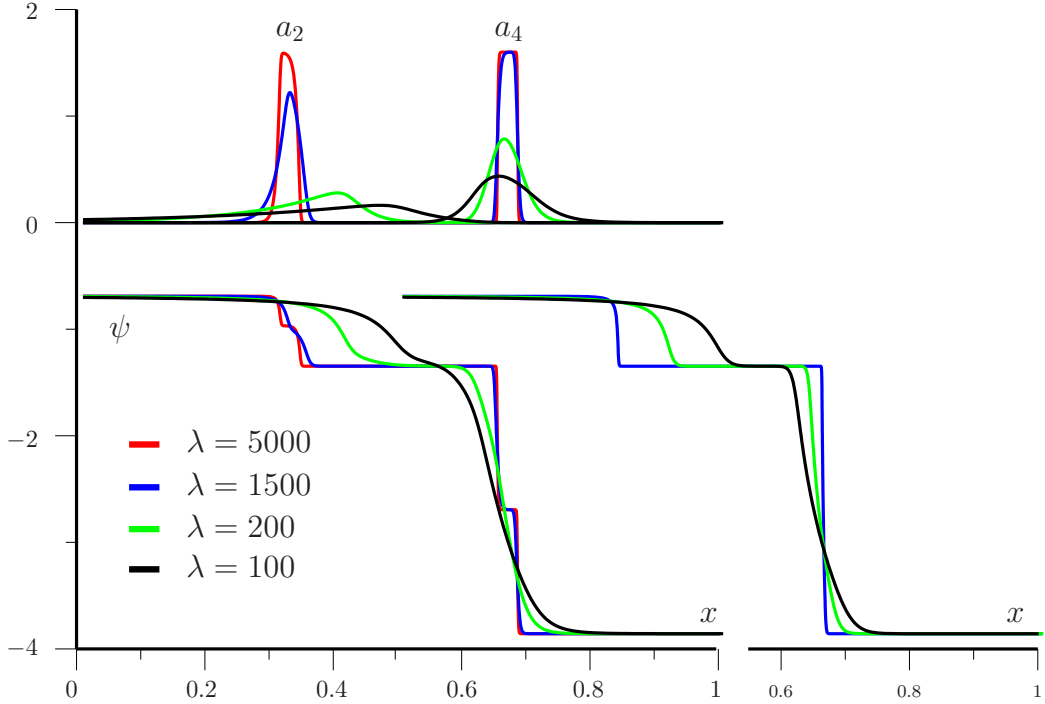}\\  [-3mm]
\caption{The distributions of the acidity $\psi(x)$ and the conductivity $\sigma(x)$ at $M_1=0.5$, $M_2=0.05$, $M_3=0.5$, $M_4=0.05$,  $M_5=0.5$.
 $\lambda=100$,  $U_0=2.382$ ($I_*=24.797  \,\,\mu\textrm{A}$,  $U_*=5.934  \,\,\textrm{V}$);
 $\lambda=200$,  $U_0=2.589$ ($I_*=49.594  \,\,\mu\textrm{A}$,  $U_*=12.994  \,\,\textrm{V}$);
 $\lambda=1500$, $U_0=2.808$ ($I_*=371.953 \,\,\mu\textrm{A}$,  $U_*=106.268 \,\,\textrm{V}$);
 $\lambda=5000$, $U_0=2.830$ ($I_*=1239.843\,\,\mu\textrm{A}$,  $U_*=357.190\,\,\textrm{V}$). See table~\ref{tab:table4}
}
\label{Ris11}
\end{figure}

Here we demonstrate the creation of $\textrm{pH}$-gradient with the help of the tree-component mixture $\textrm{His-His}$ ($a_1$),  $\textrm{His}$ ($a_3$),  $\textrm{Tyr-Arg}$ ($a_5$). This $\textrm{pH}$-gradient is used for identification of two mixture components: $\textrm{His-Gly}$ ($a_2$),  $\beta-\textrm{Ala-His}$ ($a_4$). From the mathematical point of view, we have again the five-component mixture. To simulate the identification of the components we choose
\begin{equation*}
M_1=0.5, \quad M_2=0.05, \quad M_3=0.5, \quad  M_4=0.05, \quad M_5=0.5.
\end{equation*}
Other word, we assume that the concentrations of the three components of the five-component mixture $a_1$, $a_3$, and $a_5$ are more than the concentration $a_2$, $a_4$. On Fig.~\ref{Ris11} (right) the $\textrm{pH}$-gradient in the tree-component mixture is shown at different values of the parameter $\lambda$. In the case when mixture contains $a_2$ and $a_4$ components the original $\textrm{pH}$-gradient is deformed weakly, because the concentrations $a_2$ and $a_4$ are small. At $\lambda=100$ and $\lambda=200$ the acidity $\psi(x)$ is smoothly varying function.

However, the substances $a_2$ and $a_4$ are weakly separated and are poorly identified. On the contrary, at large values of the parameter $\lambda$ the acidity $\psi(x)$ is  almost a step function, but there is good separation of components.

\subsubsection{Ten-component mixture}\label{App:ZhS-2.2.3}

Here we demonstrate the example of the ten-component mixture consisting of the amino acids that have little specific conductivity. Physicochemical parameters of the component are given in the Tab.~\ref{tab:table5}. With the help of component $a_1$ and $a_{10}$ we model electrode solutions, setting the concentration of these components to be large enough:
\begin{equation*}
M_1=0.6, \quad M_2=M_3=\dots=M_9=0.1, \quad  M_{10}=0.6.
\end{equation*}
Other components of the mixture create the $\textrm{pH}$-gradient in the electrophoretic chamber.

\begin{table}[H]
\caption{\label{tab:table5} Parameters of amino acids for ten-component and twelve-component mixture. Mobility of ions $\mu_i$ are taken from \cite{Zhukova-bib5}}

\begin{ruledtabular}
\begin{tabular}{lccccccccc}
& $n=10$ & $n=12$ & $\textrm{pKb}_i$ & $\textrm{pKa}_i$ & $\textrm{pI}_i$ & $\psi_i$ & $\delta_i$  & $\mu_i$& $ M_i$\\
\hline
Thr  & ($a_1$)    &   ($a_1$)    & $2.63$ & $10.40$ & $6.515$ & $3.661$ & $1227.35$  & $2.73$ & $0.6$ \\
Pro  & ($a_2$)    &   ($a_2$)    & $2.00$ & $10.60$ & $6.300$ & $3.108$ & $1509.97$  & $2.19$ & $0.1$ \\
Ala  & ($a_3$)    &   ($a_3$)    & $2.34$ & $9.87 $ & $6.105$ & $2.924$ & $1315.13$  & $2.87$ & $0.1$ \\
Iso  &            &   ($a_4$)    & $2.36$ & $9.68 $ & $6.020$ & $2.889$ & $1458.71$  & $2.44$ & $0.01$ \\
Leu  & ($a_4$)    &   ($a_5$)    & $2.36$ & $9.60 $ & $5.980$ & $2.567$ & $1599.44$  & $2.17$ & $0.1$ \\
Val  & ($a_5$)    &   ($a_6$)    & $2.32$ & $9.62 $ & $5.970$ & $2.509$ & $1068.98$  & $2.29$ & $0.1$ \\
Phe  & ($a_6$)    &   ($a_7$)    & $2.58$ & $9.24 $ & $5.910$ & $2.371$ & $2233.41$  & $2.63$ & $0.1$ \\
Trp  &            &   ($a_8$)    & $2.38$ & $9.39 $ & $5.885$ & $2.348$ & $2084.34$  & $2.44$ & $0.01$ \\
Met  & ($a_7$)    &   ($a_9$)    & $2.28$ & $9.21 $ & $5.745$ & $2.256$ & $2285.44$  & $2.51$ & $0.1$ \\
Ser  & ($a_{8}$)  &   ($a_{10}$) & $2.31$ & $9.15 $ & $5.730$ & $2.060$ & $2910.51$  & $3.07$ & $0.1$ \\
Gln  & ($a_{9}$)  &   ($a_{11}$) & $2.17$ & $9.13 $ & $5.650$ & $1.611$ & $9976.31$  & $2.78$ & $0.1$ \\
Asn  & ($a_{10}$) &   ($a_{12}$) & $2.02$ & $8.80 $ & $5.410$ & $1.116$ & $3836.80$  & $2.63$ & $0.6$ \\
\end{tabular}
\end{ruledtabular}
%\footnotetext[1]{$B_i=10^{-\textrm{pKb}_i}$.}
%\footnotetext[2]{$A_i=10^{-\textrm{pKa}_i}$.}
\end{table}

\begin{figure}[H]
\centering
\includegraphics[scale=0.8]{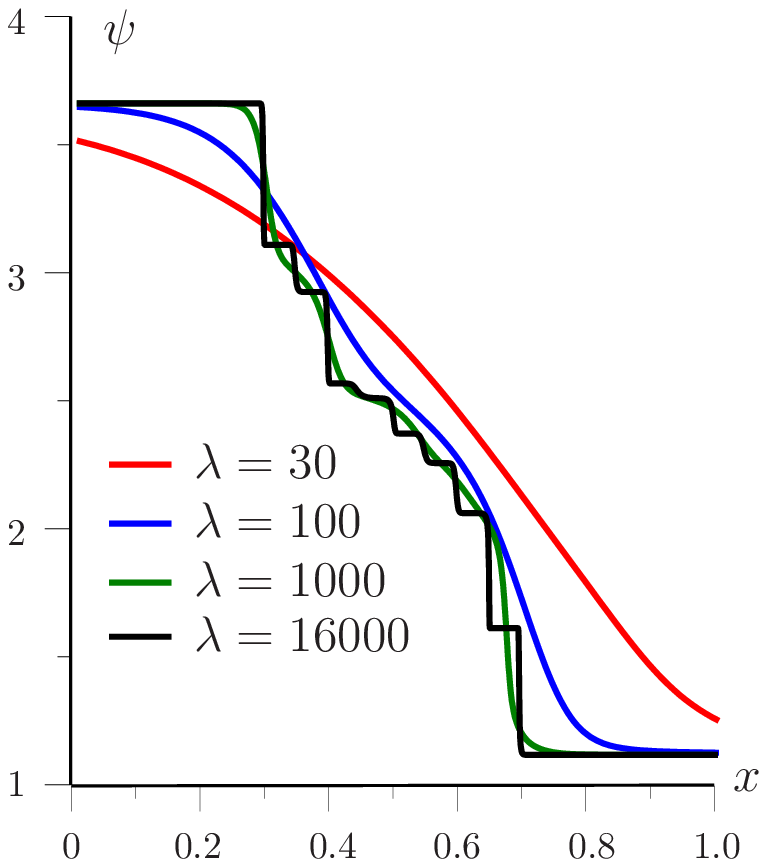}
\includegraphics[scale=0.8]{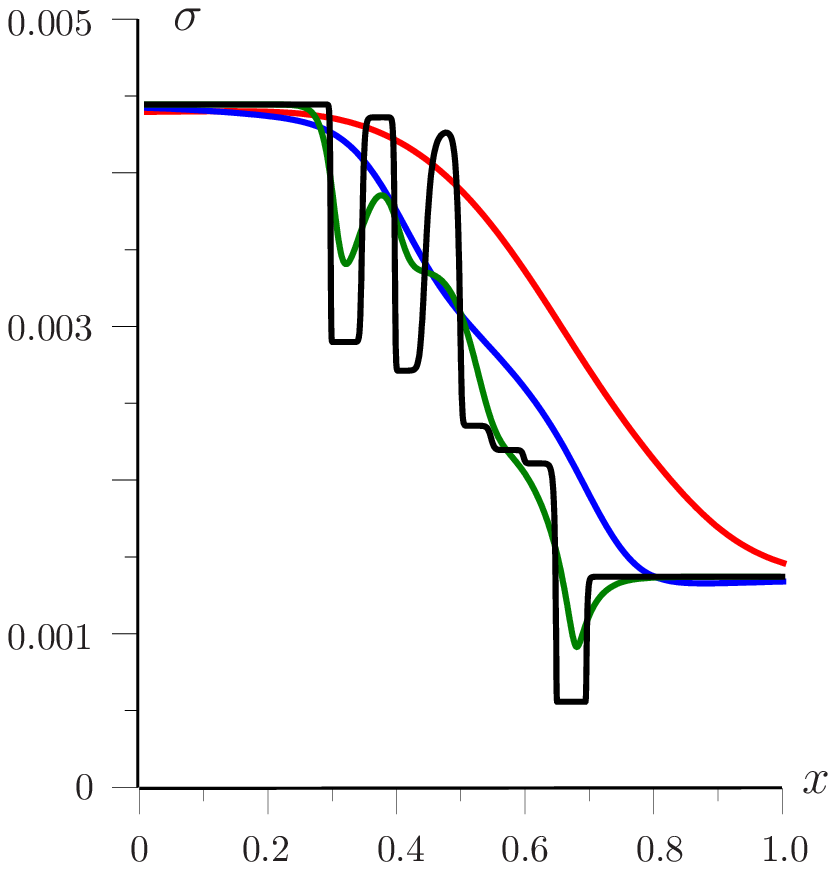}\\ [-3mm]
\caption{The distributions of the acidity $\psi(x)$ and the conductivity $\sigma(x)$. $\lambda=30$,     $U_0=333.849$ ($I_*=7.439     \,\,\mu\textrm{A}$,  $U_*=252.851         \,\,\textrm{V}$);
$\lambda=100$,    $U_0=413.893$ ($I_*=24.797    \,\,\mu\textrm{A}$,  $U_*=1045.001        \,\,\textrm{V}$);
$\lambda=1000$,   $U_0=460.696$ ($I_*=247.968   \,\,\mu\textrm{A}$,  $U_*=11631.985       \,\,\textrm{V}$);
$\lambda=16000$,  $U_0=499.085$ ($I_*=3967.497  \,\,\mu\textrm{A}$,   $U_*=201620.590     \,\,\textrm{V}$);
$\lambda=16700$,  $U_0=499.234$ ($I_*=4141.075  \,\,\mu\textrm{A}$,   $U_*=210504.319     \,\,\textrm{V}$).
See table~\ref{tab:table5} for ten-component mixture}
\label{Ris12}
\end{figure}

\begin{figure}[H]
\centering
\includegraphics[scale=0.8]{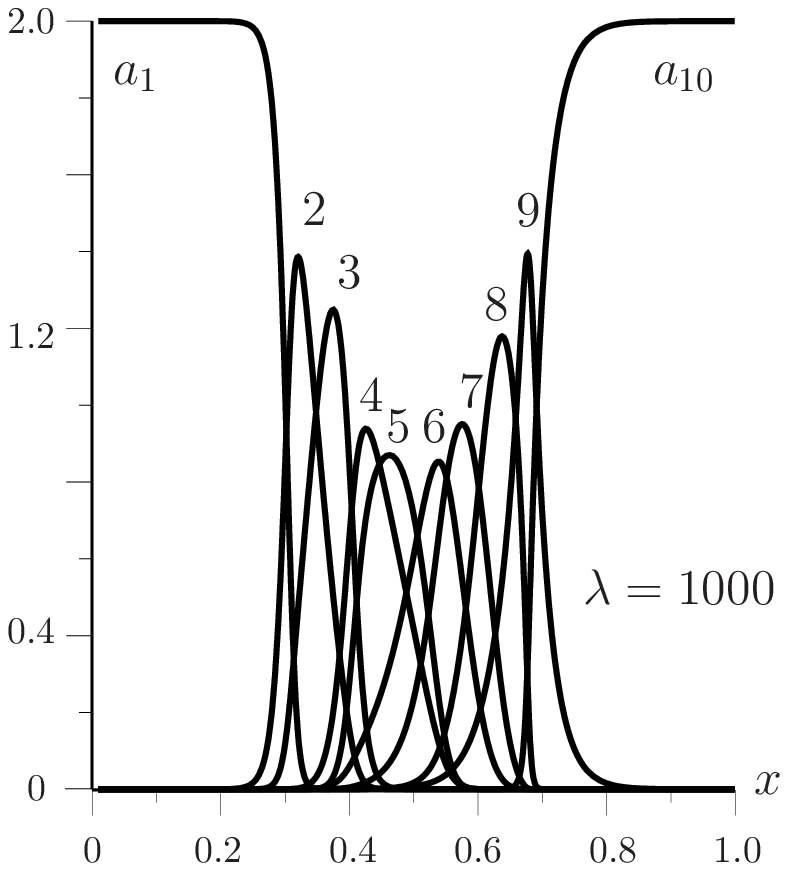}
\includegraphics[scale=0.8]{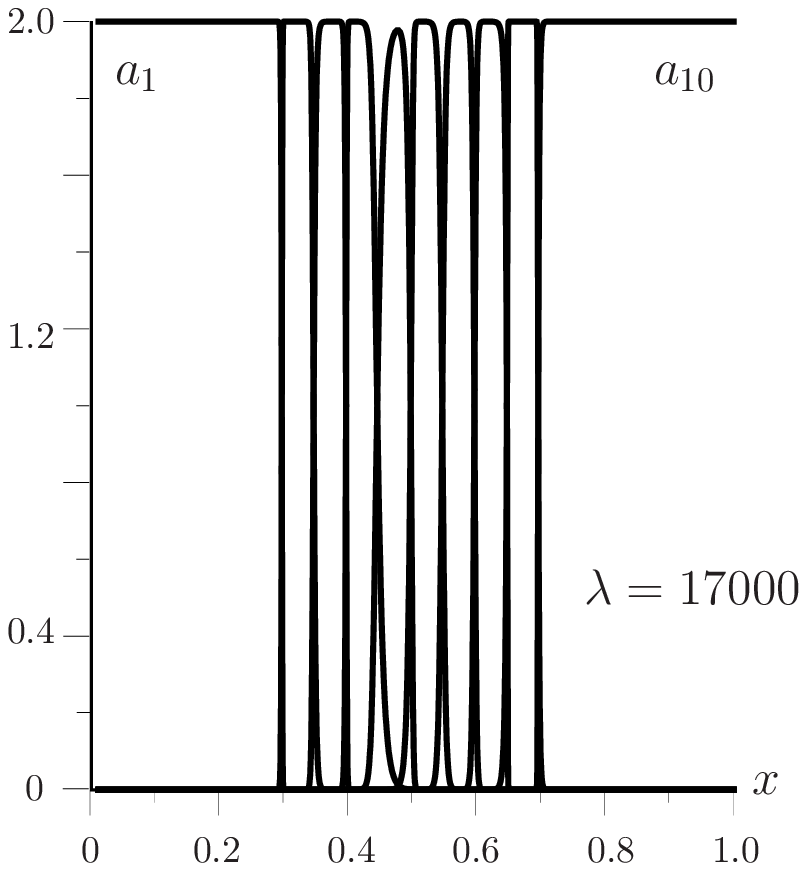}\\
\caption{The distributions of concentrations.
See table~\ref{tab:table5} for ten-component mixture}
\label{Ris13}
\end{figure}

On Fig.~\ref{Ris12} it is clearly seen that the $\textrm{pH}$-gradient becomes a step function at $\lambda=16000$. At $\lambda=17000$ we have a generalized solution (\ref{ZhSeq-E3}) of the problem (\ref{ZhS-eq7})--(\ref{ZhS-eq11}) (see Fig.~\ref{Ris13}). Unfortunately, in practice, such a result is not reachable as electric current, voltage, and Joule heat are very large (Joule heat is approximately $800$\,$\textrm{W}$).

\begin{figure}[H]
\centering
\includegraphics[scale=0.85]{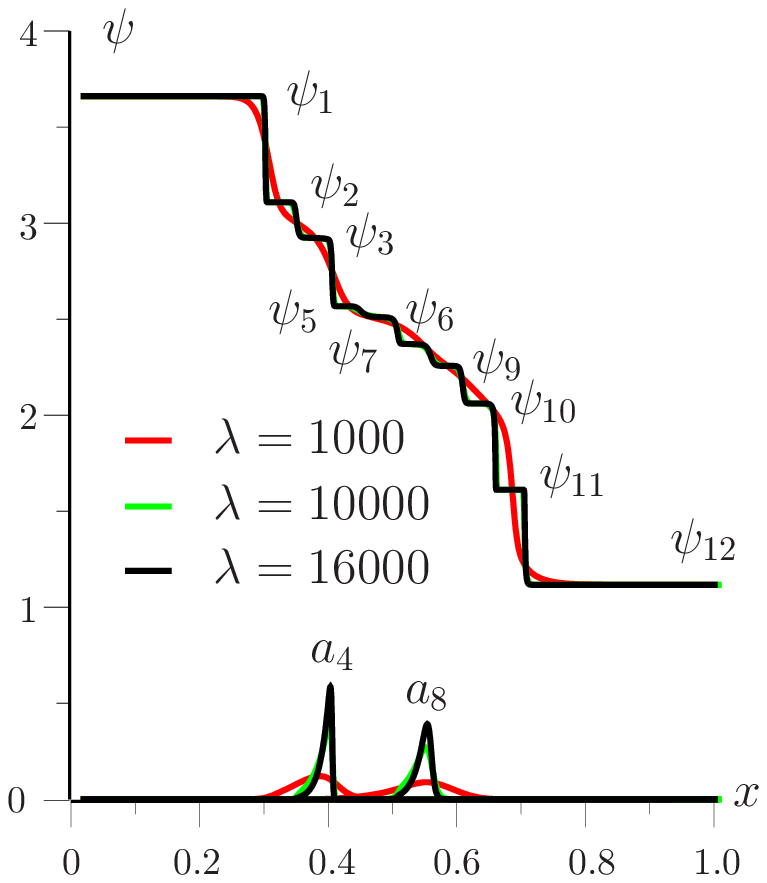}
\includegraphics[scale=0.85]{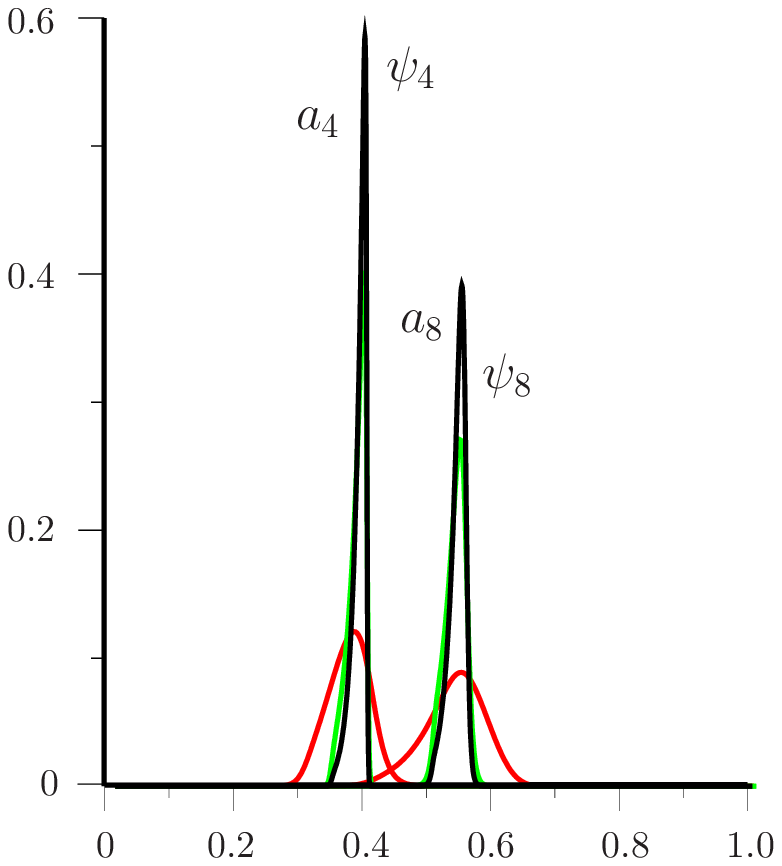}\\
\caption{The distributions of the acidity $\psi(x)$ and the conductivity $\sigma(x)$. $\lambda=1000$,   $U_0=454.85$ ($I_*=247.968   \,\,\mu\textrm{A}$,  $U_*=11484.381        \,\,\textrm{V}$);
$\lambda=10000$,  $U_0=490.57$ ($I_*=2479.686  \,\,\mu\textrm{A}$,  $U_*=123862.925       \,\,\textrm{V}$);
$\lambda=16000$,  $U_0=492.70$ ($I_*=3967.497  \,\,\mu\textrm{A}$,  $U_*=199041.174       \,\,\textrm{V}$). See table~\ref{tab:table5}  for twelve-component mixture}
\label{Ris14}
\end{figure}

On Fig.~\ref{Ris14}  the separation of the two components in the given  $\textrm{pH}$-gradient is shown. At $\lambda=17000$ it is theoretically possible to be a good separation of the $a_4$ and $a_8$ components of the twelve-component  mixture. However, in the experiment it is difficult to realize the value of the parameter $\lambda$ more than the $\lambda=1000$. In this case, although there is a separation of component $a_4$ and $a_8$, but this fractionation is not sufficiently clear.

\section{Conclusion}\label{ZhS-13}

One of the most interesting results is the fact that at $\lambda=\infty$ a generalized solution of the original problem is occurred (see (\ref{ZhSeq-E3}) and \cite{Part1}). At moderate values of the parameter $\lambda$ approximation of a weak solution is actually the asymptotic of the original problem solution. Confirmation of this fact is a good coincidence of the weak solution of the problem and the numerical solution of the problem.
In the Sec.~\ref{App:ZhS-2.2.2} another remarkable result is described.
In fact, a new variant of the IEF process is considered.
Usual IEF process is occurred to the following scheme:
1) the stable $\textrm{pH}$-gradient is created, 2) the samples of the substances are separated in given $\textrm{pH}$-gradient. The results of the
Sec.~\ref{App:ZhS-2.2.2} show that the process of separation and identification can be realized simultaneously (see Appendix~\ref{App:ZhS-2.2.2} and Figs.~\ref{Ris11}, \ref{Ris14}). Suppose we need to identify the presence of certain amino acids in the mixture, the concentration of which is small enough.
To create a gradient one can add to this mixture some number of amino acids with great concentration. The main $\textrm{pH}$-gradient is created by substances with large concentration.
Substances with a small concentration also participate in creating $\textrm{pH}$-gradient, weakly distorting the main gradient. In the stationary state the mixture is separated into individual components. Note that this method of separation opens opportunity to use IEF process in microdevice. In particular, the electric current and voltage are small enough: electric current $I_*\approx 370 \,\,\mu\textrm{A}$, voltage $U_*\approx 100\,\,\textrm{V}$, and Joule heat is approximately $37 \,\,\textrm{mW}$ (see Fig.~\ref{Ris11} at $\lambda=1500$).
In more detail the process of separation will be described in the \cite{Part3} which gives the solution of non-stationary problem.

\begin{acknowledgments}
This research is partially supported by Russian Foundation for Basic Research (grants 10-05-00646 and 10-01-00452),
Ministry of Education and Science of the Russian Federation
(programme `Development of the research potential of the high school', contracts  14.A18.21.0873,  8832 and grant 1.5139.2011).
The authors are grateful to N.\,M.~Zhukova for reviewing the translated text into English.

\end{acknowledgments}

%%%%% end zhuk %%%%%%%

%\newpage %Just because of unusual number of tables stacked at end

\setlength{\bibsep}{4.0pt}

\end{document}